\documentstyle[epsf,aps,12pt]{revtex}

\begin{document}

\draft

\def\beq{\begin{equation}}
\def\eeq{\end{equation}}
\def\ber{\begin{eqnarray}}
\def\eer{\end{eqnarray}}
\def\l {\label}
\def\rN{\rho_N}
\def\rr{(\rho_N/\rho_0)}
\def\xirr{\xi\rho_N/\rho_0}
\def\ben{\begin{enumerate}}
\def\een{\end{enumerate}}
\def\bei{\begin{itemize}}
\def\eei{\end{itemize}}
\def\et{{\it et al}}  
\def\i{\item}
\def\beq{\begin{equation}}
\def\eeq{\end{equation}}
\def\l{\label}
\def\ber{\begin{eqnarray}}
\def\eer{\end{eqnarray}}
\def\v#1{\vec{#1}}
\def\sd{$\sigma$ }
\def\sv{\langle\sigma\rangle_{vac} }
\def\s{\sigma}
\def\o{\omega}
\def\od{$\omega$ } 
\def\ov{\langle\omega\rangle_{vac} }
\def\rN{\rho}
\def\er{\xi\rN}
\def\bs{\bar{\psi}}
\def\NaP{Nielsen and P\`{a}tkos }
\def\gc{g_\chi}
\def\F{F_\pi}
\def\Fs{F^*_\pi}
\def\gs{g_{\sigma NN}}
\def\go{g_{\omega NN}}
\def\gr{g_{\rho NN}}
\def\gp{g_{\pi NN}/2M}
\def\fr{f_{\rho NN}/2M}
\def\gss{g_{\sigma NN}^*}
\def\gos{g_{\omega NN}^*}
\def\grs{g_{\rho NN}^*}
\def\gps{(g_{\pi NN}/2M)^*}
\def\frs{(f_{\rho NN}/2M)^*}

\title{Nuclear Matter Studies with Density-dependent
Meson-Nucleon Coupling Constants}  
\author{M. K. Banerjee$^{1,2}$ and J. A. Tjon$^{1,3}$}
\maketitle
\begin{center}
\end{center}
$^1$Department of Physics, University of Maryland, College
Park, Maryland 20742, USA\\
$^2$Instute for Kernphysik, Forschungszentrum J\"{u}lich, 52425 J\"{u}lich,
Germany\\  
$^3$Institute for Theoretical Physics, University of Utrecht,
 3508 TA Utrecht, the Netherlands
\date{\today}
\maketitle
\begin{abstract}

Due to the internal structure of the nucleon, we should expect, in general, 
that the effective meson nucleon parameters may change in   nuclear medium.  We
study such changes by using a chiral confining model of the nucleon.  We  
use  density-dependent  masses for  all mesons except the pion.  Within a
Dirac-Brueckner analysis, based on the relativistic covariant structure of the
NN amplitude, we show that the  effect of such a density dependence in the NN
interaction on the  saturation properties of nuclear matter, while not large,
is  quite  significant. Due to the density dependence  of the $g_{\sigma NN}$,
as predicted by the chiral confining  model, we find, in particular, a looping
behavior of  the  binding energy at saturation as  a function of the saturation
density. A simple model is described, which exhibits looping and which is shown
to be mainly caused by the  presence  of a peak in the density dependence of
the  medium modified  $\sigma N$ coupling constant at low density.

The effect of density dependence of the coupling constants and the meson
masses  tends to improve the results for $E/A$ and density   of  nuclear matter
at saturation. From the present study we see that the relationship between
binding energy and saturation density may not be as universal as found in
nonrelativistic studies and that more model dependence is exhibited once medium
modifications of the basic nuclear interactions are considered. 
\end{abstract}
\hspace{0.6cm}{PACS numbers: 21.65.+f, 14.20.-c }


\section{Introduction} 

Since a nucleon is not a point object, but
has structure, it must undergo changes when placed inside a nucleus. Among
other properties, the meson-nucleon coupling constants may change. If this
happens, it should affect the NN force. These effects have to be small.
Otherwise, traditional nuclear physics, where one uses free-space two-body
force would have failed badly. But even small changes in the NN force may have
noticeable effect on   the  properties of nuclear matter. The purpose of this
paper is to investigate possible changes of meson nucleon coupling constants
due to the quark structure of the nucleon and their ultimate   effects on the
saturation properties, such as the  density $\rho_0$ and the binding energy  
$-E/A$   of nuclear matter.

The NN force in nuclear matter may become density dependent due to a wide
variety of reasons. Any time one eliminates some degrees of freedom the
resulting effective interaction becomes density dependent, the Brueckner
$G$-matrix being the  most widely known example. Another recent example is the
work of Li \et~\cite{LI} on the effective interaction to be used in a mean
field calculation which reproduces the results of a Brueckner-Hartree-Fock
calculation.

The density dependence studied here involves excitations  of N$^*$ ($I=J=1/2$
resonances)  degrees of freedom.\footnote{The nuclear matter is an isoscalar
and isotropic medium. It cannot change the spin or the isospin  of a nucleon.
It is also translationally invariant. But, as we will see later, it can still
produce internal excitations in a nucleon.}  Ultimately we are interested in
the change of the NN force in the medium. In relativistic nuclear physics the
most important forces are mediated by a scalar-isoscalar field \sd with
$m_\s\simeq 600$ MeV and a vector-isoscalar field \od with $m_\o=783$ MeV. Thus
the force range is short and the neighboring nucleons which can alter the
internal structure of the interacting nucleons must also be close. Due to the
Exclusion principle and short-range correlations,  the effective density of the
polarizing nucleons, denoted as $\xi\rho$,  is less than half  the normal
nuclear density. It is quite possible that the $N^*$ excitations produced by
the neighboring nucleons may be treated perturbatively. However, an earlier
study of this problem~\cite{MKB} established that a fairly large number of
resonances contribute. The mean field approach, which generates the
eigen-combination of N and $N^*$s as the lowest state in the field due to the
neighboring nucleons is a more expeditious way of calculating the effect.

Our quantitative studies are based on the following strategy. We describe the
structure of the nucleon with a model called the chiral confining model
(CCM)~\cite{KIM,CCM}. Specifically, we use the TOY~\cite{CCM} version of this
model. The role of nuclear matter is simulated with baths of external \sd and
\od fields, the vacuum values being $\sv=-F_\pi=-93$ MeV and $\ov=0$,
respectively. The nucleon structure problem is solved in the presence of these
bath fields. Then various properties of the nucleon, including meson nucleon
coupling constants, are calculated for ranges of values of the two bath fields.

The fact that the \sd and the \od fields are, in turn, produced by the nucleons
through field-dependent coupling constants allows us to obtain density
dependences of the \sd and the \od fields   by solving   appropriate nonlinear
self-consistency equations. Once this has been done, we can use the known field
dependences of various properties of the nucleon to obtain their density
dependences.

We indicate the density-dependent values of various properties of the nucleon
with a  star.  Thus $\gs$ denotes the free space \sd NN coupling constant and
$\gss$ the same quantity in nuclear matter. The latter is always dependent on
$\rN$, the nuclear density. We may note that   the properties of nuclear matter
depend principally on $g^*_{\sigma NN}$, $g^*_{\omega NN}$, $g^*_{\pi NN}$,
$g^*_{\rho NN}$ and $f^*_{\rho NN}$, where the last is the $\rho$ Pauli
coupling coefficient.

A necessary input is the field-dependence   of the exchanged meson
masses\footnote{The possible importance of density dependence of meson masses
was first stressed by Brown and his collaborators~\cite{GEB}}. We use the
simple model that the meson masses are linearly dependent on the \sd field,
keeping the pion mass fixed. \footnote{Because of charge conjugation symmetry 
the masses of  mesons,   considered here, cannot   depend linearly on \od. The
dependence must be      quadratic or of higher even power.} The   coefficient
of linear dependence of \sd and \od masses on the \sd field is chosen to give 
$m^*/m=0.92$ at normal nuclear density.

The changes, as found in the CCM,  clearly will have effect on the properties
of nuclear matter. To study this we  adopt the relativistic many-body
approach~{\cite{THM,M,TJON2}} . Specifically,  we carry out  a relativistic
Dirac-Brueckner calculation of the properties of nuclear matter  using the 
one-boson-exchange model of Ref.~\cite{fleis}.  For the quasi-potential version
of this model the full Dirac  structure of the NN amplitude in free space has
been  constructed~\cite{TJON1}. The resulting so-called IA2 representation can
be used to determine the saturation properties of nuclear matter~\cite{TJON2}. 
Modifying the free space T-matrix to also include Pauli-blocking and
introducing our density-dependent meson-nucleon coupling constants,  
self-consistent relativistic Dirac-Brueckner calculations were performed in the
manner of Ref.~\cite{TJON2} for a range of $\xi\rho$ the effective density of
the polarizing nucleons.

Our main results are that \begin{itemize}
\i[] (i) the effects of the density-dependent meson nucleon coupling
constants, arising out of the quark structure of the nucleons, on the
saturation density and $-E/A$ at saturation are small but not negligible,
\i[] (ii)  and they do tend to improve the results.
\end{itemize}

The next section contain brief introduction to the Chiral Confining Model (CCM)
and its TOY version. Section 3 describes the calculation of nucleon properties
as functions of bath $\sigma$ and $\omega$ fields. These results are used in
the next section to extract density dependences of nucleon properties. Section
5 describes the relativistic treatment of nuclear matter using the
Dirac-Brueckner approach. The last section presents the main results, 
discussions of these results and concluding remarks.

\section{The Chiral Confining Model} 

An early attempt at extracting density dependence of nucleon properties using
CCM is described in Ref.~\cite{MKB}. The present  paper contains two significant
improvements - inclusions of  the instanton induced interaction and  the pion
cloud contributions to $g_{\sigma NN}$, $g_{\rho NN}$ and $f_{\rho NN}$
coupling constants.  The CCM has been described in detail in
\cite{MKB,KIM,CCM}. Here we give a  brief review.

The CCM is based on the notion of color dielectric function as introduced by
Nielsen and P\`{a}tkos~\cite{NP}. By considering the average of all possible
link operators, starting from $x-\epsilon$ and ending on $x$ with the paths
completely contained in a four-dimensional hypercube of side $L$, they
introduced a color singlet, Lorentz scalar quantity  $K(x)$ and a color octet,
Lorentz vector, coarse grained gluon field $B^a_\mu$: \ber
K(x)&=&\lim_{\epsilon\rightarrow 0}\frac{1}{N_c}Tr\,\,\, \left[\langle
e^{-i\int_{x-\epsilon}^x dy\cdot A(y)}\rangle\right], \nonumber \\
B_\mu=\frac12\sum_a \lambda_a B^a_\mu&=&\lim_{\epsilon\rightarrow
0}i\frac{\partial}{\partial \epsilon}\,\,\,  \left[\langle
e^{-i\int_{x-\epsilon}^x dy\cdot A(y)}\rangle\right]. \nonumber \eer Upon
integrating out the QCD gluon fields in favor of these new collective variables
one obtains the Nielsen-P\`{a}tkos lagrangian in the form of a derivative
expansion: \beq {\cal L}_{NP}=\bs(x)[iK(x)\frac12\stackrel{\leftrightarrow}
{\partial\!\!\!/}-K(x)m_q-gB\!\!\!/(x)]\psi(x)
-\frac{K^4}{4}G^a_{\mu\nu}G^{a\,\mu\nu}+\ldots.  \l {NPL1}  \eeq  The gauge
field is $ B^a_\mu/K$ and not $B^a_\mu$. The coarse grained field tensor is 
\beq  G^a_{\mu\nu}=\partial_\mu \frac{B^a_\nu}{K}-\partial_\nu
\frac{B^a_\mu}{K} + f^{abc}\frac{B^b_\mu}{K}\frac{B^c_\nu}{K}.  \l {NPL2} \eeq
From the gluonic term one identifies  $\epsilon=K^4$ as the color dielectric
function. \NaP conjectured that \beq \langle K\rangle_{vac}=0. \l {Kvac1} \eeq
This conjecture, crucial for our model,  has been justified from the lattice
gauge point of view by Lee \et~\cite{LCB}.

A quark has ever-present interaction with the quark condensate of the
vacuum. If $\langle K\rangle_{vac}=0$, the interaction will appear to
be infinitely strong compared to  quark kinetic energy  and a quark
cannot exist in that region. It can only reside in the region where
$\langle K\rangle_{vac}\neq 0$. A color singlet quark system can
polarize the vacuum and change the value of $K$ away from zero, thus
dynamically generating the {\em bag} where the quarks can stay. 

The Nielsen-P\`{a}tkos lagrangian recognizes the existence of gluon
condensate through the vanishing of $\langle K\rangle_{vac}$. However,
the quark condensate is not manifest. Without it one cannot develop an
effective lagrangian which contains the physics of the interaction of a
quark with the quark condensate.  We deal with this problem by
conjecturing that one can integrate out  the coarse grained gluon fields
 in favor of  meson fields as new   collective variables~\cite{CCM,MKBNCD,BBCKFL}. 

It is also necessary to introduce, following Nielsen and
P\`{a}tkos~\cite{NP}, a new color singlet, Lorentz scalar field,
$\chi$, proportional  to $K$ by the equation \beq K(x)=\gc \chi(x). \l{chi}
\eeq  Being related to $K$, a purely gluonic object, the $\chi$  field
is a member of the glueball family. Hence, it is a chiral singlet, a
fact evident from the first term of the Nielsen-P\`{a}tkos lagrangian
given by Eq.~(\ref{NPL1}). Large $N_{color}$ analysis shows that it is
a hybrid field~\cite{CCM,BLC}. The result (\ref{Kvac1}) imposes the
requirement that \beq \langle \chi\rangle_{vac}=0. \l {chivac} \eeq

Retaining minimum powers of fields and their derivatives the basic
lagrangian    of the CCM has the form:
\ber
{\cal
L}_{CCM}&=&K(x)\bs(x)[i\frac12\stackrel{\leftrightarrow}{\partial\!\!\!/
}-m_q]\psi(x) \nonumber \\
&+&\bs(x)\frac{g_\pi\{\sigma(x)+i\gamma_5\v \tau\cdot\v
\pi\}+g_\omega\omega\!\!\!/+g_\rho\v \tau\cdot(\v
\rho\!\!\!/+\gamma_5\v A\!\!/_1)}{K(x)}\psi(x)\nonumber \\
&+&{\cal L}_{meson} + \frac12 \partial_\mu\chi\partial^\mu\chi -
U(\chi). 
\l {LCCM} 
\eer
 The quantities $g_\pi$, $g_\omega$ and $g_\rho$ are the quark-meson
coupling constants. The quantity $m_q$ is the current quark mass. Its
value is set at $7.5$ MeV. It contributes $17$ MeV to the $\pi$N sigma
term~\cite{CCM}. Because of its negligible role in the present work we
will neither refer to this term nor count it as a parameter in our
subsequent discussions. However, it is included in the actual numerical work.

We use the chiral invariant lagrangian of  Lee and Nieh~\cite{LN} for 
${\cal L}_{meson}$.  The lagrangian ensures that $\langle
\sigma\rangle_{vac}=-F_\pi$ and that the mesons have their respective 
physical masses when calculated at the classical ( or tree) level. The
fields $\pi$ and $\sigma$ form a $(1/2\times1/2)$ representation of
$SU_L(2)\times SU_R(2)$, while the fields $\v \rho\pm \v A_1$ form
$(1,0)$ and $(0,1)$ representations. 

 To complete the definition of ${\cal L}_{CCM}$ one must specify the
$\chi$ potential. Since nothing substantial is known about it, 
we try two simple forms:
\ber
{\rm PURE\,MASS:}\,\,& &U(\chi)=\frac12 m_\chi\chi^2, \nonumber \\
{\rm QUARTIC:}\,\,& &U(\chi)=\frac12 m_\chi\chi^2(1-\chi/\chi_0)^2. 
\l {Uchi} 
\eer 
The QUARTIC potential has two minima, the one at $\chi=0$
describes the vacuum while the other at $\chi=\chi_0$ is an `excited'
state, which for simplicity we keep  degenerate with the vacuum. 
The value of $\chi_0$, the location of the second minimum is set at
$40$ MeV. The hybrid mass, $m_\chi$ is set at $1400$ MeV. As we will
discuss later, the results of the calculations depend largely on one particular
combination of these parameters and  rather weakly on individual ones.

Any mean field calculation containing isovector fields requires using
states which do not have good isospin symmetry. To accomplish
this we use hedgehog spinors and fields, introduced first for $\pi$ and
$\sigma$ fields by Chodos and Thorn~\cite{CT} and extended to vector
meson fields by Broniowski \et~\cite{BB}. 
We closely follow the mean field analysis of Ren \et~\cite{RB}.
 
It should be emphasized that the appearence of $K$ in the denominator 
of the quark-meson interaction
term is not {\it ad hoc}. It is obtained by matching the $K$ dependence
of the four-quark interactions which arise on the one hand from ${\cal
L}_{NP}$ by integrating out the $B_\mu$ fields and on the other hand
from ${\cal L}_{CCM}$ by integrating out the meson fields~\cite{CCM,MKBQMN,Ren1}.

The presence of the factor $K$ with
$\bar{\psi}i\frac12\stackrel{\leftrightarrow}{\partial\!\!\!/}\psi$
changes the field canonically conjugate to $\psi$ from the usual
$i\psi^\dagger$ to $iK\psi^\dagger$.  As a result the quark term in all
Noether's currents carries the factor $K$.   The transformation
$\sqrt{K}\psi(x)\rightarrow\psi(x)$ makes the pair $\psi$ and 
$i\psi^\dagger$ canonical and removes the factor $K$ from all Noether's
currents. Two changes occur in ${\cal L}_{CCM}$. The free quark term no
longer has the factor $K$, while the quark meson interaction acquires
the factor  $K^2$ in place of $K$ in the denominator.

Mean field treatment of ${\cal L}_{CCM}$ with a variety of reasonable
sets of parameters  reveals~\cite{CCM,RB,Ren1}  that even   where the
quark density is large the meson fields differ only slightly from their
respective vacuum values. This suggests strongly that we introduce a
simplified  version of CCM. The simplified version, which we call  the
TOY model,   consists of fixing all meson fields at their vacuum values:
\ber
\langle\sigma\rangle_{vac}=-F_\pi, \nonumber \\
\langle\v \pi\rangle_{vac}=\langle\omega\rangle_{vac}=\langle\v
\rho\rangle_{vac}=\langle\v A_1\rangle_{vac}=0,
\eer
leaving only the quark fields and the $\chi$ field  as dynamical
variables. The Toy model lagrangian, in terms of the canonical quark
field, is given below.
\beq
{\cal
L}_{Toy}=\bs(x)[i\frac12\stackrel{\leftrightarrow}{\partial\!\!\!/}-m_q-
\frac{g_\pi F_\pi}{(\gc\chi)^2}]\psi(x)+\frac12
\partial_\mu\chi\partial^\mu\chi - U(\chi). \l {LTOY} \eeq
Not counting $m_q$, the TOY model has only two parameters, $g_\pi
F_\pi/g_\chi^2$ and $m_\chi$ in the PURE MASS version and one additional
parameter, $\chi_0$, in the QUARTIC version.
 Formation of the bag is even more transparent in the TOY model as the
constituent quark mass term, $\bar{\psi}\frac{g_\pi
F_\pi}{(\gc\chi)^2}\psi$,  becomes infinite when $\chi\rightarrow 0$.

The   unusual form  of the quark-meson interaction  with
the factor of $K^2=(\gc\chi)^2$ in the denominator is of some
importance in the present investigation. In its absence a constant
external \od field will merely shift the energy of the quark without
exciting it to higher states. Thus the nucleon will not be polarized. A
constant \sd field will excite the quark. Orthonormality of different
eigenstates of the Dirac hamiltonian ensures that $\int
d^3ru'^{\dagger}u=0$, but not the vanishing of $\int d^3r \bar{u}'u$.
The latter gives the effect of a constant external \sd field. The
presence of the factor $(\gc\chi)^2$ enables a constant external \od
field to excite the quark and enhances the ability of a constant
external \sd field to do the same.

Because of its importance it is necessary to provide some evidence for
the presence of  $(\gc\chi)^2$ in the denominator of the quark-meson
interaction term.  Fortunately, there is a verifiable consequence of
this feature.   Gauge invariance ensures that there is no such
modification of the photon-quark interaction. Thus contrary to the
conjectures of universal coupling~\cite{SAK} or current-field
identity~\cite{CFI} the isovector couplings of the photon and the
$\rho$ meson differ by this $K^{-2}$ factor. Because of this
difference, CCM predicts that the $\rho$-nucleon Pauli coupling
constant, $f_{\rho NN}$ should differ from the isovector anomalous
magnetic moment, $\kappa_{I=1}$. The predicted~\cite{RB} value is 
$f_{\rho NN}/\kappa_{I=1}=1.4$, instead of $1$ predicted by the two
conjectures. Based on dispersion theoretic analysis of $\pi N$
scattering data H\"{o}hler and Pietarinen~\cite{HP} had estimated the
ratio to be $\sim 1.75$. Our value is within the possible uncertainties
of this result~\cite{H}.\footnote{ In the MIT bag model also the photon
and the $\rho$ meson couple to the quark in different manner. Brown and
Rho had exploited this feature to explain the value of $f_{\rho
NN}/\kappa_{I=1}$. Dialing the chiral angle they fitted the value
$1.75$. Our result is approximately independent of parameters because
of the scaling property mentioned later.}

The role of the pion field on baryon properties is never negligible. 
In the TOY model one  takes account of it
perturbatively~\cite{KIM,KIM2} wherever appropriate.  In the context of
the present study such corrections are important in the calculations of
the  coupling constants of  mesons of even $G$-parity, $\gs$, $\gr$ and $\fr$.

The Toy model with perturbative pionic correction gives poor results
for the masses of $N$ and $\Delta$ and for the nucleon charge
radii~\cite{KIM,KIM2}. McGovern {\it et al}~\cite{McG} showed that
there is approximate, but  very good scaling behavior in the TOY model.
If one chooses the parameters to get the nucleon mass right the values
of most other quantities are nearly fixed. Thus the bad  results for
mass and size cannot be improved simultaneously by varying parameters.
One needs a new interaction which is short-ranged and more attractive
between singlet-singlet quark pairs than between triplet-triplet pairs. 
The magnetic one-gluon exchange interaction~\cite{DRGG} fits the
requirement exactly. However, the physics of exchange of a color octet 
tower of gluons is already incorporated in meson exchanges. We cannot
include it again. 
 
We solve the problem by including the instantanton induced 't Hooft
interaction~\cite{Belavin,thft3}.   
In the two flavor case of $u$ and $d$ quarks, 
the 't Hooft interaction generates interactions  between quark pairs 
in flavor
antisymmetric state only and the interaction is attractive and zero-ranged. 
So it fits our requirement without duplicating any mechanism already
included. Using Shuryak's~\cite{Shuryak} description of the vacuum as
an instanton liquid the strength of the 't Hooft interaction  can be
estimated in QCD, but not  in CCM~\cite{KIM}. So we have to treat it as
a free parameter. We  fix it by fitting the $N-\Delta$ mass splitting
with the 't Hooft interaction together with one-pion exchange
contribution~\cite{KIM}. 

The method of calculation with the TOY model in the absence of external
\sd and \od has been described in detail in Ref.~\cite{KIM} and will
not be repeated here. The changes due to the presence of external
fields  are straightforward~\cite{MKB}. Going back to ${\cal L}_{CCM}$,
described by Eq.~(\ref{LCCM}), we interpret $\sigma$ and $\omega_0$ as
the bath fields.  The vacuum expectation values of all other fields,
$\v \pi$, $\omega_i$ ($i=1,2,3$), $\v \rho_\mu$ and $\v A_{1\mu}$ are
zero   because nuclear matter has good isospin and  good  parity. The
quark field and the $\chi$ field continue to be the only dynamical
variables in the extended TOY model lagrangian.  The new energy
functional of a nucleon in a \sd-\od bath is given below.
\ber
E&=& N_c\int d^3r u^\dagger(\v r)\left[-i\v \alpha\cdot\v \nabla -
\frac{g_\pi\gamma_0\sigma-g_\omega\omega}{(\gc\chi(\v r))^2}\right]u(\v
r)\nonumber \\
&+&\int d^3r\left[\frac12(\v \nabla\chi(\v r))^2+ U(\chi(\v r))\right],\nonumber \\
&-&2C_s\int d^3r{\cal Z}(r)(G^2(r)+F^2(r)).
 \l {NMTOY} 
 \eer 
where the valence spinor, $u(\v r)$ is  
\beq
u(\v r)=\left(\begin{array}{c}
G(r) \\
i\vec{\sigma}\cdot\hat{r}F(r)
\end{array}
\right)\zeta, 
\label {spinor} 
\eeq
and $\zeta$ is a 2-component Pauli spinor.

The last line in Eq.~(\ref{NMTOY}) represents the contribution of the
instanton induced 't Hooft interaction with a regulating function, 
${\cal Z}(r)$, which cuts off the interaction at a distance scale of $0.25$ fm. The details may be found in 
Refs.~\cite{KIM,KIM2}.

 Note that the external fields are constant in space and time.
The vacuum value of the \sd field is $-F_\pi$. Nuclear matter generates
\sd field with positive sign thus reducing the magnitude of the net
\sd. It is customary to represent the effect as a modification of the
pion decay constant:\beq \sigma=-F_\pi+\sigma_{nmatter}=-F_\pi^*. \l
{fpistar} \eeq The \od in Eq.~(\ref{NMTOY}) is the time component of
vector \od field. The static nuclear matter distribution can generate
only the time component.  However, a nucleon moving with velocity $\v
v$ in nuclear matter does see a space component $\v \omega=-\v
v\omega/\sqrt{1-v^2}$. We ignore the role of this term.

\section{Field Dependence of Nucleon Properties}
\l {FDNP}
 
We pick values of $\Fs$ in the range $63$ MeV to $93$ MeV in steps 
of $6$ MeV and values of \od in the range $0$ to $40$ MeV in 
steps of $8$ MeV. 
For each pair of $\Fs$ and \od we find the spinor and the $\chi$ field
which will make the energy functional stationary. The condition of
stationarity yields coupled nonlinear equations. Ref.~\cite{KIM}
describes the method of solving these equations. Once the solutions are
obtained we can calculate the desired properties of the nucleon in the
\sd-\od field bath.

We introduce the dimensionless variables $x$ and $y$ to describe, in
units of $m_\pi$, the bath fields \sd and   \od:   \ber
\sigma-\langle\sigma\rangle_{vac}&=&\F-\Fs=xm_\pi. \l {xdef} \\
\omega&=&ym_\pi, \l {ydef} \eer We fit the results of every physical
quantity $Q$ of interest with the quadratic form:
\beq 
Q(x,y)=a[1+bx+cy+dx^2+fxy+gy^2]=aF^Q(x,y). 
\l {FQdef} 
\eeq 
\begin{table}[tp]
 \caption{ Values of quantities a, b, c, d, f and g for the  PUREMASS
case. The quantities under the column a are the free space values.}
\label{Tab:Pxy}
\begin{center}
\begin{tabular}{||l|c|c|c|c|c|c||}\hline\hline
 Quantity        & a    & b        & c     & d     & f      & g \\ \hline
 $(g_{\pi NN}/2M)^{*}$& 0.83 & 1.72    & -0.78 & 3.51    & -5.71  &
0.13  \\ \hline
$g_{\omega NN}^{*}$  &11.02  &  1.10    & -1.60 &   2.44  & -2.90  & 3.22   \\ \hline
$g_{\sigma NN}^{(q)*}$  &2.34  &   3.14   & -5.25 &   3.34  & -21.77 &
6.99   \\ \hline\hline
\end{tabular}
\end{center}
\end{table}

\begin{table}[tp]
\caption{ Values of quantities a, b, c, d, f and g for the QUARTIC
case. The quantities under the column a are the free space values.}
\label{Tab:Qxy}
\begin{center}
\begin{tabular}{||l|c|c|c|c|c|c||}\hline\hline
                                
     Quantity    & a    &    b     &   c   &     d   &   f    &   g    \\ 
\hline
 $(g_{\pi NN}/2M)^{*}$& 0.92  &1.62   & -0.49&    3.55 &  -3.36&  0.045   \\ \hline
$g_{\omega NN}^{*}$ &13.72 & 0.74& -0.93&   1.23 &   -1.18&  1.45 \\ \hline
$g_{\sigma NN}^{(q)*}$  &4.92    & 1.67&-2.13 & -2.39 &   -6.65&    
1.72 \\ \hline\hline
\end{tabular}
\end{center}
\end{table}
The results of our fit are shown in Tables~\ref{Tab:Pxy} and
\ref{Tab:Qxy} for the PUREMASS and QUARTIC case respectively.
The quantity $a$ is the value of $Q$ in vacuum and $F^Q(x,y)$ is the
quadratic polynomial in square brackets.
The   three coupling constants of special interest are $g_{\sigma
NN}^{(q)*}$, $\gos$ and $\gps$.  The $\s$-nucleon coupling constant
carries a superscript $(q)$ to indicate that   only the direct coupling
of the \sd field to  valence quarks is included in $g_{\sigma
NN}^{(q)*}$. The full coupling constant also includes coupling to the
pion field produced by the valence quarks. The \od and the $\pi$ fields
have odd G-parity. Hence their source currents  couple to $3$ or higher
odd powers of the $\pi$ field. Since the $\pi$ field is weak inside the
nucleon, the higher power contributions are not important and have been
omitted in the present calculation.  The  superscript $(q)$ is
redundant  for $\gos$ and $\gps$  and have been  omitted.

The parts of the meson source current which show the coupling to quarks
may be read off  Eq.~(\ref{LCCM}) describing ${\cal L}_{CCM}$. We list them below:
\ber
j^{(q)}_\sigma&=&g_\pi\frac{1}{K(x)}, \l {jsq} \\
j^{(q)}_\omega&=&g_\omega\frac{\gamma_0}{K(x)}, \l {joq} \\
j^{(q)}_{\pi,\alpha}&=&g_\pi\frac{i\gamma_5 \tau_\alpha}
{K(x)}, \l {jpq} \\
j^{(q,\mu)}_{\rho,\alpha}&=&g_\rho\frac{\gamma^\mu\tau_\alpha/2
}{K(x)}. 
\l {jrq} 
\eer
The meson-nucleon coupling constants are defined in the usual way. The
expressions listed below are for $ \v p\,\,\rightarrow 0$.
\ber
\langle N(\vec(p))\mid j^{(q)}_\sigma(0)\mid N(\vec(p))\rangle &=& \gs^{{(q)}} , \l {defgs} \\
\langle N(\vec(p))\mid j^{(q)}_\omega(0)\mid N(\vec(p))\rangle &=& \go  , \l {defgo} \\
\langle proton(\vec(p))\uparrow\mid j^{(q,\mu)}_{\rho,3} \mid
proton(\vec(p))\uparrow\rangle &=&\frac12\gr^{{(q)}},
\l {defgr} \\
\langle proton(\vec(p))\uparrow\mid \frac12\int 
d^3 \v r\ [\v r\times \v j_{\rho,3}^{(q)}]\  
\mid proton(\vec(p))\uparrow\rangle &=&\mu_{\rho NN}^{(q)}, 
\l {deffr} \\
\langle proton(\vec(p)')\uparrow\mid j^{(q)}_{\pi,3}(0)\mid proton(\vec(p))
\uparrow\rangle &=& (\v p'-\v p)_z \, \gp. 
\l {defgp} 
\eer 
The quantity $\mu_{\rho NN}$ appearing in Eq.~(\ref{deffr}) is the
$\rho$-magnetic moment of the nucleon:
\beq 
\mu_{\rho NN}=(g_{\rho NN}+f_{\rho NN})/2M. 
\l {rhomagdef}
\eeq

We list below the results from CCM for the contributions to the various
coupling constants coming from direct coupling of   meson fields   to the quarks. 
\begin{eqnarray}
g_{\sigma NN}^{(q)*} &=& g_\pi\int d^3r[G^2(r)-F^2(r)]/K^2(r), \label{gsq} \\
(g_{\pi NN}/2M)^* &=& g_\pi\frac{10}{9}\int d^3r rG(r)F(r)/K^2(r), \label{gpiq} \\
g_{\omega NN}^*&=& 3 g_\omega \int d^3r[G^2(r)+F^2(r)]/K^2(r) , \label{gomegaq} \\
g^{(q)*}_{\rho NN}&=& g_\rho \int d^3r[G^2(r)+F^2(r)]/K^2(r),\label{grhoq}  \\
\mu_{\rho NN}^{(q) *}&=&g_\rho\frac{10}{9}\int d^3r rG(r)F(r)/K^2(r). \label{rhomag}
\end{eqnarray} 
To obtain density-dependent, or equivalently field-dependent, coupling
constants one must use $K$ and spinor components $G$ and  $F$  obtained
in the presence of the bath fields.
The stars should be removed for  free space values of the coupling
constants  obtained with $K$, $G$ and  $F$ appropriate for free space.

The nucleon mass term which appears as a divisor for $g_{\pi NN}$ and
for the $\rho$ magnetic coupling stands for the experimental value of
the mass, {\it i.e.,} $939$ MeV. It is used to  define dimensionless
coupling constants from the integrals of dimensions of length which
appear in Eqs.~(\ref{gpiq}) and (\ref{rhomag}). Our studies measures
the density dependence of the quantities represented by the integrals.  

 A model of hadrons cannot anticipate what effective hadronic
lagrangian its results will be used for. Hence, to remove possible
misunderstanding we have stated in Eq.~(\ref{defgp}) the precise
definition of the quantity $\gp$ as it appears in our work.   The
expression on the left hand side involves the pion source current,
usually well-defined   in most quark based  model of the nucleon. It
is  also well-defined  in any effective hadronic lagrangian.   If the
effective lagrangian  uses pseudoscalar $\pi$N coupling the quantity
$g_{\pi NN}$ is the dimensionless coupling constant of the theory. If,
on the other hand, it uses derivative coupling then $\gp$ is the
coupling constant of the theory, having the dimensions of length. Note,
again, that $M$ in the denominator of $\gp$ for the derivative coupling
theory is a fixed number by definition, usually $939$ MeV.  

The mesons with even G-parity, \sd and $\rho$, also couple to the pion
cloud,   the contributions  from which  depend essentially quadratically
on $\gps$. Hence we write:
\begin{eqnarray}
g_{\sigma NN}^*&=&g_{\sigma NN}^{(q)*} +\zeta_\sigma(g_{\pi NN}/2M)^{*2},
\label {gsd} \\
g_{\rho NN}^*&=&g_{\rho NN}^{(q)*}+\zeta_g(g_{\pi NN}/2M)^{*2}, 
\label {grd} \\
\mu_{\rho NN}^*&=&\mu_{\rho NN}^{(q)*}+\zeta_f(g_{\pi NN}/2M)^{*2}. 
\label {gfd}
\end{eqnarray}  
The parameters, $\zeta_\sigma$, $\zeta_g$ and $\zeta_f$,
represent the strengths of the pion cloud contributions. The
quantities  $\zeta_\sigma$ and  $\zeta_g$ have the dimensions of mass
squared , while $\zeta_f$ has the dimensions of mass.
We choose reasonable values of $\gs$, $\gr$ and $\fr$ in free space and
use the results of the TOY model in free space  for $\gs^{(q)}$,
$\gr^{(q)}$, $\mu_{\rho NN}^{(q)}$ and $\gp$ to fix the coefficients
$\zeta_\sigma$, $\zeta_g$ and $\zeta_f$. These values and the relevant
parameters are listed in either  Table~\ref{parafile} or in the 
table caption.

We will see later in section~\ref{DBA} that the free space meson nucleon 
coupling constants used
here are not exactly the same as in the one boson exchange potential used in
the Dirac-Brueckner calculations. However, the differences are small and are
not expected to affect the qualitative purpose of this paper.

As stated in the introduction a necessary input is the field-dependence 
of the exchanged meson masses. We use the simple model that the meson
masses are linearly dependent on $x$\cite{GEB}, keeping the pion mass
fixed.   We assign the same $x$-dependence to both \sd and \od masses. 
\beq m^*=m(1-b_{meson}x)=m\phi(x). \label{mesmass} \eeq The function
$\phi(x)$ has been introduced so that the rest of the discussion is not
specific to the simple linear dependence used in the numerical work.
The   coefficient $b_{meson}$ is chosen to give  $m^*/m=0.92$ at
normal nuclear density. Its values are listed in Table~\ref{parafile}.
\begin{table}[tp]
\label {parafile}
\caption{List of parameters and coefficients. Both the PUREMASS and
QUARTIC cases use $m_{\s}=650$ MeV, $m_{\o}=783$ MeV, $m_\rho=770$ MeV,
$m_\pi=140$ MeV, $\gs=8.0$, $\gr=5.0$, $\mu_{\rho NN}=2.45$ and
$m^{*}/m=0.92$ at normal nuclear density. The numbers in the table are all in pion mass
units. The quantity  $\gp$ has the dimensions of length and is
expressed in units of $m_\pi^{-1}$. The quantities $\go$, $\gs^{(q)}$,
$\gr^{(q)}$ and $b_{meson}$ are dimensionless. The quantity $\zeta_f$ has
the dimensions of mass and is expressed in units of $m_\pi$. The
quantities $\zeta_g$ and $\zeta_f$ have dimensions of mass squared and are
expressed in units of $m_\pi^2$.}
\begin{center}
\begin{tabular}{|l|l|l|}\hline
Quantity & PUREMASS & QUARTIC \\ \hline
$\gp$ & 0.8349 & 0.9241 \\ \hline
$\go$ & 11.02 & 13.72 \\ \hline
$\gs^{(q)}$ & 2.389 & 4.919  \\ \hline
$\zeta_\sigma$ & 8.049 & 3.608  \\ \hline
$\gr^{(q)}$ & 2.448 & 3.048  \\ \hline
$\zeta_g$ & 3.661 & 2.286  \\ \hline
$\mu_{\rho NN}^{(q)}$ & 0.8349 & 0.9241  \\ \hline
$\zeta_f$ & 2.317 & 1.787  \\ \hline
$b_{meson}$ & 0.4769 & 0.4622  \\ \hline
\end{tabular}
\end{center}
\end{table}

\section{Density Dependence}

The first step in converting the information about field dependence of
the coupling constants to density dependence is to obtain the density
dependence of the fields, \sd and \od, or equivalently, $x$ and $y$, 
themselves. This is done through the following two basic self-consistency equations:
\ber
x&=&\frac{\gss}{m_\s^{*2}}\xi\rN, \label{xdensdep1}\\
y&=&\frac{\gos}{m_\o^{*2}}\xi\rN, \label{ydensdep1}
\eer
where the various quantities have been defined by Eqs.~(\ref{FQdef})
and  (\ref{mesmass}).

\begin{figure}[tbp]
\epsfxsize=4.8in
\epsfysize=2.5in
\epsffile{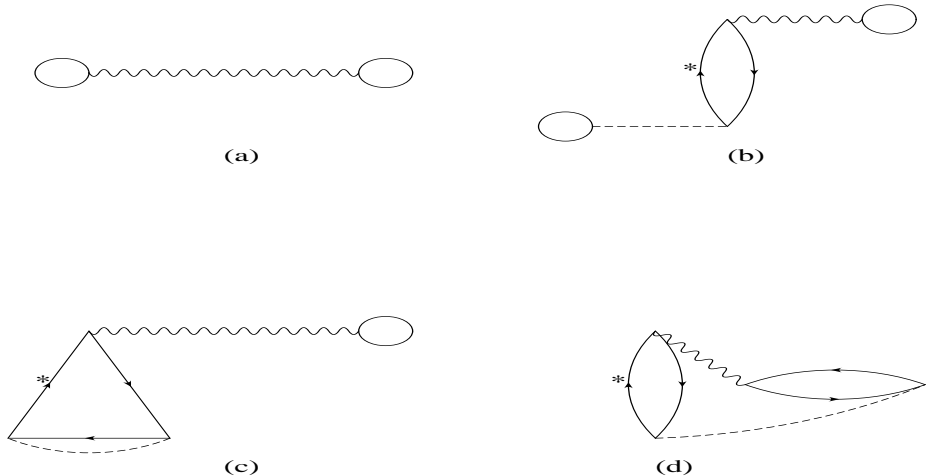}
\caption{The Fig. (a) represents interaction between two Fermi sea
nucleons interacting via a G-matrix. In Fig. (b) a third nucleon excites
one of the interacting nucleons into a resonance state, leaving a hole
in the Fermi sea. The resonance propagator is shown as an upward line
with a star.  Fig. (c) is the result of exchange between the third nucleon
and the left interacting nucleon. Recall the convention that a line
starting and ending at exactly the same time represents a hole line.
Fig. (d) represents the exchange between the third nucleon and the
right interacting nucleon. }
\label {Gold}
\end{figure}

 The factor $\xi$ represents the reduction of the effective density
of nucleon available to polarize a given interacting pair. The
reduction is the combined effect of the Exclusion principle and
short-range correlations among nucleons.   An examination of the
illustrative Goldstone graphs of Fig.~\ref{Gold} may be useful to
understand the origin of this factor. 
Besides the diagrams shown in the figure there are others obtained by
interchanging the roles of the right and the left interacting nucleons.
  There are also diagrams where a fourth nucleon excites the  
interacting nucleons to a resonance state, generating   two
resonance-two hole states. The two resonance states interact and fall
back into the hole states.  All these are examples of single
interaction of the neighboring nucleons with the interacting pair.
These   may be repeated  and summed to generate field effects in a
space containing  the  nucleon and  $I=J=1/2$ resonance states as
degrees of freedom.  If resonances were not included the procedure will
generate contributions to the usual average field field of a many body
theory where only the nucleon degree of freedom is counted.  The mean
field of  CCM   describes approximately the multidimensional average
field described here.   

 Inclusion of all exchange diagrams enforces the   Exclusion principle,
which ensures that only two other nucleons can come close to the two
interacting nucleons. The latter are mostly     within a  short
distance. This consideration alone  makes $\xi=1/2$. Repulsive short
range correlations make $\xi$ even smaller.  We expect that
$0\leq\xi\leq 0.5$.  A full self-consistent many body treatment
should determine this factor. But for the present preliminary study we
take it to be an unknown parameter confined to the range $0\leq\xi\leq 0.5$.

In Eq.~(\ref{xdensdep1}) one should use the scalar density. This can be done
only through a full many body self-consistent calculation. We simplify our work
by using the vector density $\rho$. This simplifying strategy  allows us to
factor the problem of extracting the density dependence of the coupling
constants from the rest of the many body problem.  We estimate the error to be
$\sim \langle \v p\,^2\rangle/4M^2\simeq 0.05$. The strategy is probably  
justifiable in a preliminary study of the main question studied in this paper.

The Eqs.~(\ref{xdensdep1}) and  (\ref{ydensdep1})   may be rewritten as:
\beq
x=\{\Sigma^{(q)}F^\s(x,y)+\Sigma^{\pi}(F^\pi(x,y))^2\}\frac{1}{\phi^2(x)
}\,\xi\frac{\rN}{\rho_0}, 
\label {xdensdep2}
\eeq
where the combinations $\Sigma^{(q)}$ and $\Sigma^{\pi}$ are 
\beq
\Sigma^{(q)}=\frac{g^{(q)}_{\s NN}}{m_\s^2}\,\rho_0\,\,\,{\rm
and}\,\,\,\Sigma^{\pi}=\zeta_\s(\gp)^2\frac{\rho_0}{m_\s^2}.
\label{xdensdep3} 
\eeq 
The quantity $\rho_0$ is the normal nuclear
density. The quantities $\Sigma^{\s}$ and $\Sigma^{\pi}$ would be the
\sd fields produced by nuclear matter due to direct \sd-quark coupling
and due to  \sd-pion cloud coupling, respectively, if the coupling
constants did not change with density. The $x$, $y$ dependent
quantities were defined by Eqs.~(\ref{FQdef}) and  (\ref{mesmass}).
 
 For future convenience we introduce the dimensionless effective
density $\bar{\rho}$:
\beq 
\bar{\rho}=\xi\frac{\rN}{\rho_0}. 
\label{rhobar} 
\eeq
In a similar manner we write 
\beq
y=\Omega\frac{F^{\o}(x,y)}{\phi^2(x)}\,\bar{\rho}, 
\label {ydensdep2}
\eeq 
where 
\beq
\Omega=\frac{g_{\o NN}}{m_\o^2}\,\rho_0, 
\label{ydensdep3}
\eeq 
would be
the \od field produced by nuclear matter if the coupling constant did
not change with density.

We solve the two coupled nonlinear Eqs.~(\ref{xdensdep2}) and (\ref{ydensdep2})
for a range of  $\bar{\rho}$ giving us the fields $x$ and $y$ as functions of
$\bar{\rho}$. Using the results and Eqs.~(\ref{FQdef}) we obtain  numerical
values of the five coupling constants, $\gss$, $\gos$, $\gps$, $\grs$, 
$\mu_{\rho NN}^*$  and $m^*/m$ as functions of $\bar{\rho}$.

We remind the reader that we do not use the absolute coupling constants from
the CCM calculation in the Dirac-Brueckner treatment of nuclear matter. We use
only the results for the density dependences of the ratios such as 
$g^*_{\sigma NN}/g_{\sigma NN}$.

\begin{figure}[tbp]
\hspace{0.5in}
\epsfxsize=4.8in
\epsfysize=4in
\epsffile{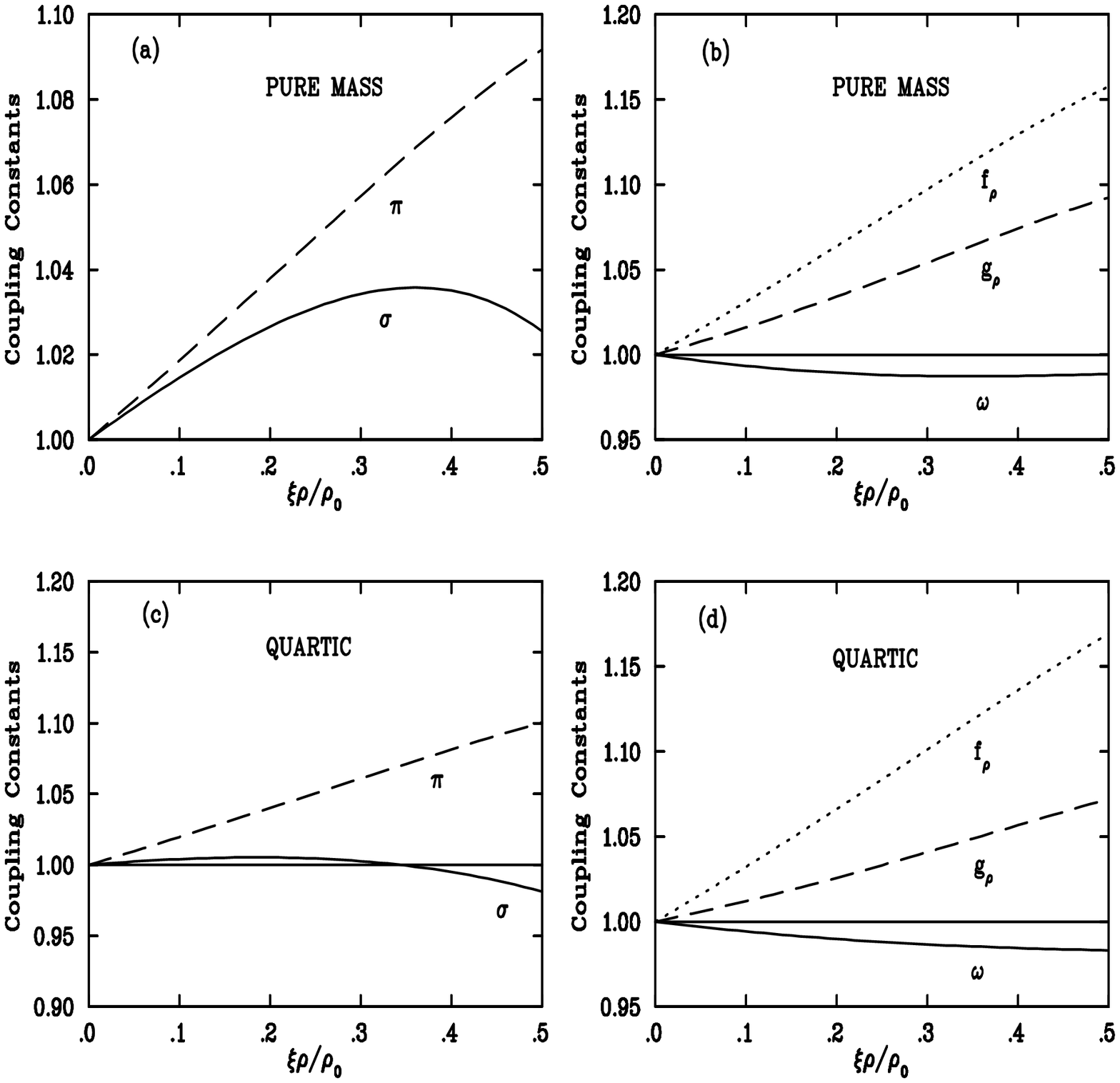}
\caption{ Plots of    $\gss/\gs$ labelled with \sd,   $\gps/(\gp)$
labelled with $\pi$, $m^*/m$,   $\gos/\go$ labelled with $\omega$,
$\grs/\gr$ labelled with $g_\rho$ and $\mu_{\rho NN}^*/(\mu_{\rho NN})$
labelled with $f_\rho$ as functions of $\xirr$. Results for both  the
PUREMASS  and the QUARTIC forms for $U(\chi)$ are presented. All
relevant  parameters are listed in Table~\protect{\ref{parafile}} in
section~\protect{\ref{FDNP}}.}
\label {coupling}
\end{figure}

These ratios are plotted as functions of  $\bar{\rho}=\xi\rho/\rho_0$
in Fig.~\ref{coupling}.
To make it convenient to use these results   in a relativisitic
Dirac-Brueckner-Hartree-Fock calculation we represent the density dependences
in  the form of a $[4,5]$ rational function: 
\ber
\frac{g^*}{g}=1+\frac{\sum_{\ell=1,4}\alpha^g_\ell\bar{\rho}^\ell}
{1+\sum_{n=1,5}\beta^g_n \bar{\rho}^n}, 
\label{gratfn} \\
\frac{m^*}{m}=1+\frac{\sum_{\ell=1,4}\alpha^m_\ell\bar{\rho}^\ell}
{1+\sum_{n=1,5}\beta^m_n\bar{\rho}^n}. 
\label{mratfn} 
\eer 
The coefficients $\alpha_\ell$ and $\beta_n$ for the five coupling
constants and the meson mass ratio are obtained by least square fitting
and are listed in Tables~\ref{pratco} and \ref{qratco}.

\newpage
\begin{table}[tp] 
\caption{ Coefficients of rational functions defined in
Eqs.~\protect{(\ref{gratfn})} and \protect{(\ref{mratfn})} obtained
with PURE MASS $U(\chi)$. Columns 2 through 6 are for the ratios of
coupling constants, while the last column is for $m^*/m$. Rows labelled
1 through 4 are the numerator polynomial and the remaining rows are for
the denominator polynomial.  All    parameters relevant to the
calculations are listed in Table~\protect{\ref{parafile}} in
section~\protect{\ref{FDNP}}.}      
\begin{center}   
\begin{tabular}{|l |l|l|l|l|l|l|} \hline
&$\frac{\gss}{\gs}$&$\frac{\gos}{\go}$&$\frac{\gps}{\gp}$&$\frac{\grs}
{\gr}$&$\frac{\mu_{\rho NN}^*}{\mu_{\rho NN}}$&$m^*/m$ \\ \hline
   \,1&    0.1553&   -0.0779&    0.1817 &    0.1491&    0.3015&   -0.0877 \\ \hline
   \,2&   -0.4452&    0.2303&   -0.4327&   -0.2797&   -0.7174&    0.1535\\ \hline
   \,3&    0.4843&   -0.2198&    0.4802&    0.2101&    0.7951&   -0.1481\\ \hline
   \,4&   -0.2853&    0.0484&   -0.2026&   -0.0700&   -0.3352&    0.0295\\ \hline
   \,5&   -2.3792&   -1.3216&   -2.6652&   -2.6050&   -2.7329&   -2.1338\\ \hline
   \,6&    2.8749&    0.3065&    3.6722&    3.1574&    3.8297&    2.6547\\ \hline
   \,7&   -1.7181&    1.0527&   -2.6344&   -1.8743&   -2.8004&   -1.6752\\ \hline
   \,8&    0.6688&   -0.6230&    1.2151&    0.5764&    1.2733&    0.9508\\ \hline
   \,9&   -0.0517&    0.1177&   -0.1483&   -0.0486&   -0.1415&   -0.1372\\ \hline 
\end{tabular}
\end{center}
\label {pratco}
\end{table}
\begin{table} 
\caption{ Coefficients of rational functions defined in
Eqs.~\protect{(\ref{gratfn},(\ref{mratfn})} obtained with QUARTIC
$U(\chi)$. Columns 2 through 6 are for the ratios of coupling
constants, while the last column is for $m^*/m$. Rows labelled 1
through 4 are the numerator polynomial and the remaining rows are for
the denominator polynomial.  All   parameters relevant to the
calculations  are listed in Table~\protect{\ref{parafile}} in
section~\protect{\ref{FDNP}}}      
\begin{center}   
\begin{tabular}{|l |l|l|l|l|l|l|} \hline
&$\frac{\gss}{\gs}$&$\frac{\gos}{\go}$&$\frac{\gps}{\gp}$&$\frac{\grs}
{\gr}$&$\frac{\mu_{\rho NN}^*}{\mu_{\rho NN}}$&$m^*/m$ \\ \hline 
  1&    0.0522&   -0.0653&    0.1928&    0.1117&    0.3129&   -0.0852\\ \hline
   2&   -0.2078&    0.1460&   -0.3308&   -0.2173&   -0.5382&    0.1093\\ \hline
   3&    0.2027&   -0.1240&    0.2994&    0.1304&    0.4855&   -0.0842\\ \hline
   4&   -0.1174&    0.0297&   -0.1087&   -0.0230&   -0.1756&    0.0140\\ \hline
   5&   -1.9719&   -1.0893&   -1.9685&   -2.7149&   -2.0459&   -1.5207\\ \hline
   6&    2.0516&    0.2868&    2.2234&    3.1836&    2.3486&    1.4465\\ \hline
   7&   -1.0666&    0.5555&   -1.2896&   -1.9596&   -1.3931&   -0.6018\\ \hline
   8&    0.3575&   -0.2999&    0.5296&    0.5796&    0.5573&    0.2730\\ \hline
   9&   -0.0292&    0.0451&   -0.0640&   -0.0663&   -0.0601&   -0.0200\\ \hline 
\end{tabular}
 \end{center}
\label {qratco}
\end{table}

\section{Dirac-Brueckner analysis}
\label{DBA}
Modifications of the meson masses and meson-nucleon coupling
constants by the presence of the nuclear medium as discussed in
the previous sections will, in general, lead to changes in the
saturation properties of nuclear matter. This problem can be studied
within the  relativistic Dirac-Brueckner approach. To carry out such 
a study we follow the approach  described in Ref.~{\cite{TJON2}}. 
As dynamical input we use
the relativistic one-boson-exchange model of Ref.~\cite{fleis}. 
The NN T-matrix satisfies a Bethe-Salpeter equation, which is
formally given by
\beq
T = V +V S_2 T,
\l {tm}
\eeq
where V is the NN interaction and $S_2$ is the free 2-nucleon
Green function.
In this study we use in particular the quasi-potential version of it.
The interaction $V$ is assumed to be given by the exchange of $\pi$, $\sigma$,
$\rho$, $\omega$, $\delta$ and $\eta$ mesons. 
In particular, in the model a
derivative pion and $\eta$ meson coupling has been assumed.
For our study we choose the coupling parameters of
interaction A, which gives a reasonable fit to the NN scattering 
phase shifts{\cite{hummel}}. To regulate the behaviour at high
momenta, a monopole form factor $\Lambda^2/(k^2-\Lambda^2)$ has
been introduced at each meson-nucleon vertex. A cutoff mass of
$\Lambda=1150$ MeV has been taken.
In Table~\ref{amodel} the meson parameters of the model A are listed.
\begin{table}[tp] 
\caption{ Meson-nucleon coupling constants $g$ and meson masses
$m$ of the one-boson-exchange model A}
\begin{center}   
\begin{tabular}{|l |l|l|l|l|l|l|l|} \hline
&${\pi}$ & ${\sigma}$ & ${\omega}$ & ${\rho^v}$ & 
$ \rho^t/\rho^v$ &
$ {\delta} $ & $ {\eta} $ \\ \hline
$g^2/4\pi$  & 14.2 & 7.6 & 11.0 & 0.43 & 6.8 & 0.75 & 3.09 \\
m (MeV) & 139 & 570 &783 & 763 & 763  & 960 & 548 \\ \hline
\end{tabular}
 \end{center}
\label {amodel}
\end{table}

According to the CCM model the free space meson parameters will be modified in
the presence of the nuclear medium. Within the one-boson-exchange model the
medium modifications found in the previous section can be simply  implemented
by replacing for each meson $\phi$ the coupling constant $g_{\phi NN}$ and the
mass $m$ in the one-boson-exchange interaction $V$ by the  corresponding
$g^*_{mNN}$ and $m^*$  given by Eqs.~(\ref{gratfn}) and (\ref{mratfn}).
Moreover, the Pauli-blocking due to the medium has to be accounted for  in
Eq.~(\ref{tm}), leading to the Bethe-Brueckner-Goldstone equations for the
G-matrix. This is done by replacing the Green function in Eq.~(\ref{tm}) \beq
S_2 \rightarrow S(p_1) S(p_2) Q \l {pbg} \eeq where $S(p_n)$ are the
medium-modified nucleon propagators and $Q$ the Pauli-blocking operator, which
projects out in the intermediate states nucleon momenta inside the Fermi
sphere.  Introducing the  baryonic current $B=\rho\,u$, with $\rho$ being the
density  and $u$ the unit vector $u=(1,{\v 0})$  in the nuclear-matter frame,
due to Lorentz covariance we may write the nucleon self-energy contribution in
terms of the 3 invariants  $\Sigma^\alpha$  \beq \Sigma(p) =\Sigma^s -\Sigma^0
\gamma.u-\Sigma^v \gamma.p_\perp, \l {self} \eeq where $p_\perp= p - (p.u) u$.
The medium-modified nucleon propagator can be approximated near the dressed
nucleon pole as \beq S(p) =  [ E^* \gamma_0 - {\v p} {\v \gamma} - M^* ]^{-1}
\l {prop} \eeq with  \ber \label{mst} M^* = (M+\Sigma^s)/(1+\Sigma^v)\\   
\nonumber E^*= (p_0+\Sigma^0)/(1+\Sigma^v) \eer For the Pauli-blocking operator
$Q$ an angular averaged approximation has been made. The resulting
relativisitic G-matrix equations \beq \label{gm} G= V + V S(q_1) S(q_2) Q G
\eeq with $q_n$ the momenta of the nucleons in the intermediate states, are
solved in the $NN$ CM frame after partial wave  decomposition. This is done
within a quasi-potential equation  description using the helicity basis of
positive and negative  energy spinor states corresponding to mass $M^*$.

Relativistic nuclear matter calculations require the
knowledge of the transformation properties of the G-matrix  under Lorentz
transformations. After having determined
the relativistic G-matrix in the 2-particle CM system
at a fixed matter density, we use the IA2 representation~\cite{TJON2} 
to obtain it in the nuclear matter frame.
This representation gives the complete covariant form of the
G-matrix in an unambiguous way. To reconstruct it all matrix
elements of the amplitude in the full Dirac space are 
needed. From this IA2 representation, neglecting the vacuum 
fluctuation terms, the self-energy can be readily determined.
Taking nucleon 2 to be one of the valence nucleons in the 
Fermi sphere with Fermi momentum $p_f$, we have
\beq
\label{sigm}
\Sigma(p) = -Tr_2 \int_{|p_2|<p_f} \frac{d \v{p}_2}
{(2 \pi)^3 2 E_2^{*}}  <p,p_2| G| p,p_2>,
\eeq
where nucleon 2 is in a positive energy state with
momentum $p_2$ and $Tr_2$ designates the summation over the  
spin index of this nucleon and
$E_2^*=\sqrt{ p_2^2 + M^{*2}}$.
Hence the self-energy contribution is essentially obtained 
by taking the diagonal 
matrix-elements of the G-matrix with one of the nucleons 
belonging to the filled Fermi sphere while the other one carries
a momentum $p$.

Since the G-matrix depends implicitly on the $\Sigma^\alpha$'s
through Eq.~(\ref{prop})
the calculations have to be carried out in a self-consistent way.
In doing so, the self-energy can be determined for a given
matter density $\rho$.
In particular we  have assumed that $\Sigma$ does not
vary significantly within the Fermi sphere, so that its value can 
be taken at the Fermi momentum $p_f$.
Some studies~{\cite{THM}} of the momentum dependence  have been made,
indicating that the variations in $\Sigma$ are of the order of 
10 $\%$. Hence  this is a reasonable approximation. 

The partial wave decomposed integral equations for the G-matrix 
have been solved using Gaussian quadratures. Typically 24
Gaussian points are sufficient to get an accuracy of a few
percent.  Moreover, 6 Gaussian points for each integration 
variable in Eq.~(\ref{sigm}) have been used to determine the
self-energies.
The self-consistent solutions can be found in an iterative
way.  Adopting a starting value for the $\Sigma$'s in
Eq.~(\ref{mst}), Eq.~(\ref{gm}) is solved to yield through
Eq.~(\ref{sigm}) new values for the $\Sigma$'s.
Varying subsequently the $\Sigma$'s we then can
determine the solutions of Eqs.~(\ref{gm}-\ref{sigm}) such that
the $\Sigma$'s are the same.
Once the selfconsistent self-energy solutions have been 
found the binding energy of the
ground state can readily be calculated from the energy-momentum
tensor as a function of density.
For more details of Dirac-Brueckner
calculations we refer to Refs~{\cite{THM}} and {\cite{TJON2}}.

For the various relativistic one boson interactions we find, in general, 
that the system exhibits saturation. 
As found in Ref~{\cite{TJON2}}, an exception to 
this is when one assumes that only the meson-masses drop
as $m^*/m=M^*/M$~{\cite{GEB}}. It is interesting to
note that in the case we also allow for
medium-modification of the meson coupling
constants as is predicted in our study of the chiral
confining model, saturation does occur, that is
we find that the binding energy of the ground state as a
function of the matter density $\rho$ has a minimum.
In the next section we discuss our results for the saturation 
properties of nuclear matter as predicted for the medium
modifications of the coupling constants as found in our CCM
model.

\section{Results and Discussions}
\subsection{Main Results}
As we have discussed earlier,  the coupling constants are dependent not directly
on the density $\rN$ of nuclear matter, but on the effective density,
$\xi\rN$, seen by an interacting pair in its immediate neghborhood. The Pauli
Exclusion principle ensures that $\xi<1/2$. There is certainly further
reduction of $\xi$ due to strong repulsion at short distances. Thus the  range of
$\xi$ of interest is $0\leq \xi \leq 0.5$. 

\begin{figure}[tbp]
\begin{center}
\vspace{-1.25in}
\epsfxsize=4in
\epsfysize=4in
\epsffile{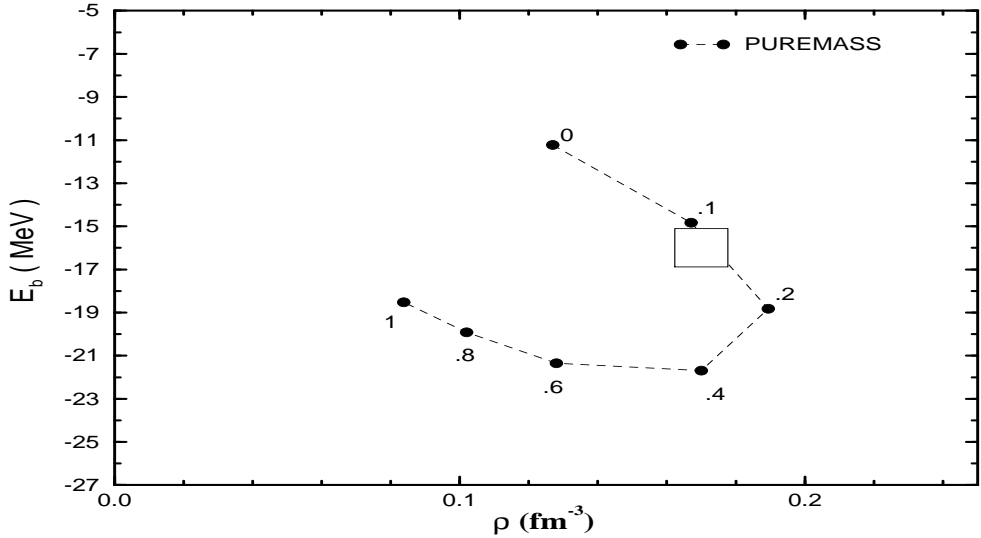}
\end{center}
\caption{Results of $E/A$ and $\rN$ at saturation for the PUREMASS case for 
a series of values of
$0 \leq \xi \leq 1.0$, shown with solid circles. 
The square box represents the empirical nuclear matter density and binding energy.
The values of $\xi$
appear next to the solid circles.  The line is for guiding the
eye only. The details of calculations are given in section~\protect{\ref{DBA}}.
Only the range $0 \leq \xi \leq 0.5$ is of physical interest.}
\label {EARP}
\end{figure}

\begin{figure}[tbp]
\begin{center}
\vspace{-1.25in}
\epsfxsize=4in
\epsfysize=4in
\epsffile{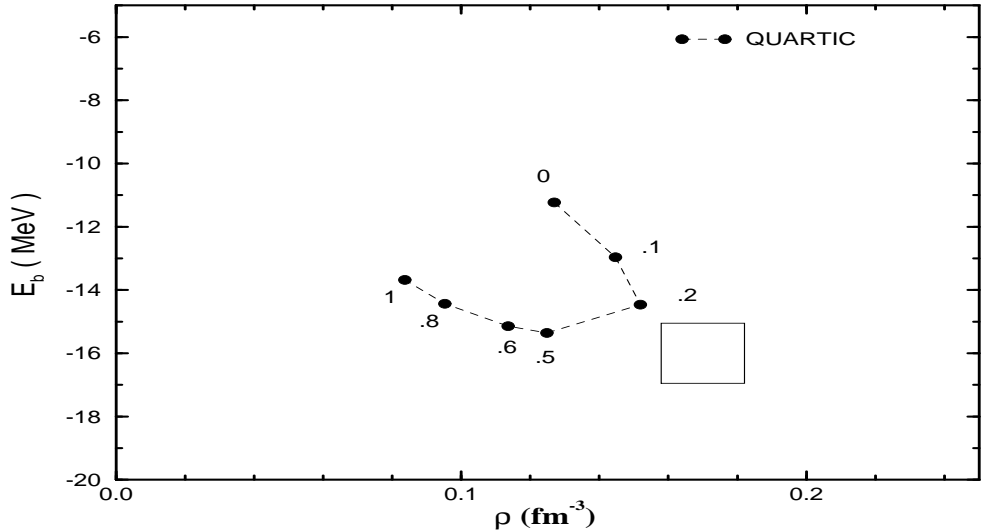}
\end{center}
\caption{Results of $E/A$ and $\rN$ at saturation for the QUARTIC case for 
a series of values of
$0 \leq \xi \leq 1.0$, shown with solid circles.  
The square box represents the empirical nuclear matter density and binding energy.
The values of $\xi$
appear next to the solid circles.  The dashed line is for guiding the
eye only. The details of calculations are given in section~\protect{\ref{DBA}}.
Only the range $0 \leq \xi \leq 0.5$ is of physical interest.}
\label {EARQ}
\end{figure}

We pick a value of $\xi$ and proceed to calculate  $E/A$ and $\rN$
at saturation using the Dirac-Brueckner approach described in 
Section~\ref{DBA}. In the calculations reported here  we have used the
one boson exchange interaction A specified in Ref.~{\cite{TJON2}}. 
During each loop of self-consistency iteration we use coupling
constants and meson masses appropriate for  $\bar{\rho}_N=\xi\rN/\rho_0$,
 defined earlier by Eq.~(\ref{rhobar}), $\rN$ being
the nuclear density at the particular  stage of the iteration. Specifically,
we use  Eqs.~(\ref{gratfn}) and (\ref{mratfn}) with the
coefficients of the rational functions listed in Tables~\ref{pratco} and
\ref{qratco}. Finally, self-consistency is achieved and we have a pair of
numbers, $E/A$  and  density $\rN$ at saturation for the given value
of $\xi$. Figures~\ref{EARP} and \ref{EARQ} show these results for the 
PUREMASS and the QUARTIC cases, respectively.

The value corresponding to $\xi=0$ represent the   Amorim and
Tjon~\cite{TJON2} result. From the figures we see that dependence of  the
nuclear saturation properties on $\xi$ is just of the right size to be of
interest in the present study. The $E/A$ vs. $\rho$ curve for the PUREMASS case
passes through the square representing experimental data for $\xi$ slightly
greater than $0.1$. For the QUARTIC case the $E/A$ vs. $\rho$ curve comes close
to the square, but does not enter it. The closest approach occurs  for
$\xi$  a little larger than $0.2$.  The results for the unphysical range,
$\frac{1}{2} <\xi \leq 1$ are included in the figures to exhibit the  curious
looping effect.

\subsection{Discussions}

In order to study the origin of the looping effect we have systematically
switched off   the density dependences of the various  coupling constants and
meson masses. In so doing, we find that  the density dependence of $\gss$ is
the main source  of the looping behavior of the saturation values of $E/A$ and
$\rN$. Qualitatively such a behaviour can be understood with a simple model
where the essential ingredient of the density dependence in the coupling
constant $\gss$ is built in as a small modification of the saturation curve.

Assuming that all meson masses and all coupling constants, except $\gss$ are
 independent of $\rho_N$, the density dependence of $E/A$ can be
modelled in the following manner: \begin{equation} E/A= (E/A)_0 + \frac12
K_0(y-1)^2-\frac34\rho\frac{\gs(0)^2} {m_\sigma^2}\left[(g_{\sigma
NN}(\rho)/g_{\sigma NN}(0))^2-1\right],  \label{App1}  \end{equation}  where,
for convenience, we have introduced the quantity  \begin{equation}
y=\rho/\rho_0,  \label{ydef1}  \end{equation}

The first two terms describe the density dependence of $E/A$ near the
saturation point when none of the coupling constants and meson masses is
density dependent. The quantity $K_0$ is the nuclear incompressibility and 
$(E/A)_0$ is the minimum value of $E/A$ in this situation.  The last term
contains the effect of the density dependence of $\gss$, calculated
perturbatively,  using, in addition,  the fact that $\overline{(\vec{p}_1-
\vec{p}_2)^2}\ll m_\sigma^2$ {\it i.e.,} the average squared relative momentum
of two Fermi sea nucleons is much smaller than the square of the \sd meson mass.

\begin{figure}[tbp]
\begin{center}
\epsfxsize=3in
\epsfysize=3in
\epsffile{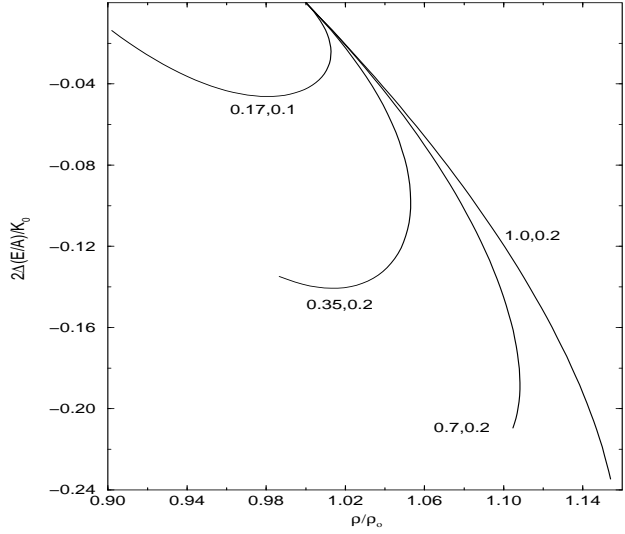}
\end{center}
\caption{Plots of $2\Delta(E/A)/(280 MeV)$ vs. the saturation density  
$\rho/\rho_0$ for $K_0=280$MeV and $(y_1,\,\eta)=
(0.17,\,0.1),\,(0.35,\,0.2),\,(0.7,\,0.2)$, and $(1.0,\,0.2)$.}
\label{loop}
\end{figure}

An inspection of the  two graph in the left column of Fig.~\ref{coupling} shows
that the quadratic form: \begin{equation} g_{\sigma NN}(\rho)/g_{\sigma
NN}(0)=1 + \eta\xi y(1-\frac{\xi y}{2y_1}), \label{gdens} \end{equation}
can describe these graphs quite well. The ratio $g_{\sigma NN}(\rho)/g_{\sigma
NN}(0)$ peaks at $\xi y=y_1$ and the peak value is $1+\eta y_1/2$. The values
of the various parameters used in the illustration are $K=280 MeV$,  $g_{\sigma
NN}=8.0$, $m_\sigma=630 MeV$. We consider four sets of $y_1$ and $\eta$,
namely,  $(y_1,\,\eta)= (0.17,\,0.1),\,(0.35,\,0.2),\,(0.7,\,0.2)$, and
$(1.0,\,0.2)$. The first two describe the ratios  $g_{\sigma
NN}(\rho)/g_{\sigma NN}(0)$ for the QUARTIC and the PUREMASS cases,
respectively,  in
Fig.~\ref{coupling}. The last two are used to illustrate the origin of  looping.

It is convenient to recast Eq.~(\ref{App1}) into the form: \begin{equation}
2[E/A-(E/A)_0]/K_0=(y-1)^2-  y\frac{3\rho_0}{4K_0}\frac{\gs(0)^2} {m_\sigma^2}
\left[(g_{\sigma NN}(\rho)/g_{\sigma NN}(0))^2-1\right]. \label{EbyA} 
\end{equation} Because of the presence of the third term in Eq.~(\ref{App1})
the saturation density will  change. For a given value of $\xi$ this can be 
readily determined by minimizing  the expression for $E/A$ with respect to
$y=\rho/\rho_0$. The resulting plots of the change in the binding energy at the
new saturation density vs. $\rho/\rho_0$ are shown in Fig.~\ref{loop} for
the four sets of  where $\xi$ has been varied in the region
$0.\leq\xi\leq 0.5.$

Starting from zero density, as $g_{\sigma NN}(\rho)$ increases with
$\bar{\rho}$ the quantity $E/A$ becomes more negative.  The saturation
properties reflect this through a combination of increasing saturation density,
$\rho$, and decreasing $E/A$. Eventually, decrease of  $g_{\sigma NN}(\rho)$
beyond its peak with increasing $\bar{\rho}$ and the effect of incompressibilty
take over and the saturation density stops increasing.  If this occurs in the
range $0\leq \xi\leq 0.5$,  we see  looping. If the peak of $g_{\sigma
NN}(\rho)$ occurs at a higher density, {\it i.e.,} for a higher $y_1$,  the
saturation density keeps on increasing and we see no looping for $\xi\leq
0.5$. In the present example the critical value of $y_1$ for $\eta=0.2$ appears
to be slightly higher than 0.7. Thus the occurence of a peak of $g_{\sigma
NN}(\rho)$ at not too high a value of the effective density is needed for
looping to occur for $\xi\leq 0.5$. The value of $y_1$ is lower for the
QUARTIC case than for the PUREMASS case. This explains why the loop turns for
the former case at a  density lower than that for the latter.

\subsection{Concluding Remarks} 

Due to the internal structure of the nucleon, we should, in
general, expect that the effective meson nucleon parameters
may change in   nuclear medium. 
Using a chiral confining model we have studied
the  density dependent changes of the meson-nucleon
coupling parameters.  We have also used a simple ansatz for the density
dependence of \sd and \od masses. 
Using the framework of a Dirac-Brueckner analysis we have 
found that their effect on the saturation properties of
nuclear matter can be significant.
Due to the density dependence 
of the $\gss$  as predicted by the chiral confining 
model we have  found, in particular, a looping behaviour in 
the nuclear matter binding energy at saturation density.
The looping has essentially also been verified in a simple 
model, where it is mainly caused by the 
presence of a peak in the density dependence of the 
medium modified $\sigma N$ coupling constant at a low density.
We should stress that the effect of the density dependence of the other
 quantities are not negligible. However, a qualitative understanding of the
 looping  effect can be obtained by paying attention to the 
 density dependence of $g_{\sigma NN}$ alone.

It appears that the density dependence of the coupling constants and the meson
masses produce effects which are small but interesting. In particular the
small  effect tends to improve the results for nuclear matter. From the
present study we see that the relationship between binding energy and
saturation density may not be as universal as found in nonrelativistic studies
and that more model dependence is exhibited once medium modifications of the
basic nuclear interactions are considered. We hope that these preliminary
results will encourage more detailed investigation of this particular variety
of density dependence of the NN interaction.

\section{Acknowledgements}
 This work was supported by DOE Grant DOE-FG02-93ER-40762.
Bulk of the writing and part of the work by one of us (MKB) were done while
visiting the Institute of Kernphysik, Forschungszentrum J\"{u}lich. He thanks
his colleagues there for their hospitality.
His visit was made possible by a  Alexander von Humboldt
Foundation award for senior U.S. physicist and he thanks 
the Foundation for it.


\end{document}